\documentclass[preprint,aps,showpacs,floatfix]{revtex4}
\usepackage{graphicx}
\usepackage{amsmath,amssymb,graphicx,epsfig}

\begin{document}

\title{Disordered vortex arrays in a two-dimensional condensate}

\author{Enik\H o J. M. Madarassy and Carlo F. Barenghi}

\affiliation{School of Mathematics, Newcastle University,
Newcastle--upon--Tyne, NE1 7RU, UK}

\date{\today}

\begin{abstract}
We suggest a method to create turbulence in a Bose-Einstein condensate. The method consists in, firstly, creating an ordered vortex array, and, secondly, imprinting a phase difference in different regions of the condensate. By solving numerically the two-dimensional Gross-Pitaevskii equation we show that the motion of the resulting positive and negative vortices is disordered.
\end{abstract}


\maketitle


\section{Introduction}
The study of the turbulence in superfluids - atomic Bose-Einstein condensates and liquid helium II \citep{Barenghi} - help us to understand better issues of classical Euler fluid dynamics \citep{Barenghi1}. In a superfluid the rotational motion can only take the form of vortex filaments with fixed quantized circulation. The length of these filaments is macroscopic (they can be as long as the experimental cell) but the thickness of the vortex core is microscopic (about $10^{-8}$ cm in liquid helium II). In 3D superfluid turbulence is manifested as a disordered tangle of vortex filaments. In 2D disordered motion of the vortices is a sign of turbulence. Indeed, many remarkable similarities between classical turbulence and superfluid turbulence have been noticed. For example the classical $k^{-5/3}$ Kolmogorov energy spectrum (where k is the wavenumber) was observed in helium II when agitated by rotating propellers \citep{Maurer}. The
same spectrum was apparent in the numerical simulations of Nore et al \citep{Nore} and of Tsubota and collaborators \citep{Araki,Kobayashi}. Another example is the classical $t^{-3/2}$ temporal decay of turbulence (where t is time) which was measured in helium II behind a  towed grid \citep{Smith}

\bigskip
In atomic Bose-Einstein condensates, it has been suggested that turbulence can be created by combining rotations about two axes \citep{Kobayashi1}. Decay of superfluid turbulence due to the interactions of vortices with rarefaction solitary waves was investigated by \citet{Berloff}. The decay and forced 2D turbulent flow \citep{Laval} was also described with numerical simulations. Large scale turbulence of quantized vortices was studied in superfluid $^3$He-B and $^4$He. The disadvantage of atomic Bose-Einstein condensate is that with existing techniques, condensates are rather small and contain at the most some hundreds vortices. On the other hand good visualization of individual vortices is possible. An interesting problem to study is the transformation of kinetic energy into acustic energy \citep{Nore} arising from vortex reconnections \citep{Leadbeater} and vortex acceleration \citep{Parker1}.

\bigskip
The aim of this paper is to show that the phase imprinting method \citep{Burger,Dobrek} can be used to study the formation of rotating turbulence accompanied by acoustic emissions in a simple 2-dimensional condensate.  

\section{Model and numerical technique}
To explore the density and phase profile of a rotating condensate, we solve numerically the following two-dimensional Gross-Pitaevskii equation which governs the time evolution of the complex order parameter $\psi(\mathbf{r},t)$: \\

\begin{equation}
i\hbar \frac{\partial \Psi}{\partial t} = \left[-\frac{\hbar^{2}}{2m}\nabla^{2}
+V_{tr}+g|\Psi|^{2}-\mu-\Omega L_{z}\right]\Psi,
\label{eqn:GP}
\end{equation}
where the coupling constant is $g=4\pi\hbar^{2}a/m$, $m$ is the atomic mass and $\emph{a}$ is the scattering length. The chemical potential, $\mu$, which can be initially estimated in dimensional form as: 

\begin{equation}
\mu= 2\hbar \sqrt{\frac{a}{m}},
\label{eqn:MuI}
\end{equation}
from the Thomas-Fermi solution \citep{Stringari} of Eq.~\ref{eqn:GP}, is found by the single-step integration:

\begin{equation}
\mu=-\frac{\ln \frac{\langle|\psi(t)|^{2}\rangle}{\langle|\psi(t+\Delta t)|^{2}\rangle}}{2\Delta t},
\label{eqn:Mu_1}
\end{equation}
after the initial Thomas-Fermi solution has relaxed to a time - independent solution in the harmonic trap, where $\langle...\rangle$ denotes spatial average. The trapping potential is 

\begin{equation}
       V_{tr}(x,y)=\frac{1}{2}m\omega_{\perp}^{2}\left(x^{2}+y^{2}\right),
\label{eqn:VTR}
 \end{equation}
where $\omega_{\perp}$ is the radial trap frequency. The quantity $\Omega$ is the angular frequency of rotation about the z-axis and $L_{z}$ = $i\hbar(x\partial_{y}-y\partial_{x})$ is the angular momentum operator.

\bigskip
In order to perform our analysis, we introduce the density and the velocity: $\rho(\mathbf{r},t)=|\psi(\mathbf{r},t)|^{2}$ and ${\bf v }= \hbar/m \nabla S({\bf r},t)$ and decompose the total energy, $E_{tot}$, in the following way:

\begin{equation}  
      E_{tot}=E_{kin}+E_{int}+ E_{q}+E_{tr}.
\label{eqn:TotE}
\end{equation}
where the kinetic energy, $E_{kin}$, the internal energy, $E_{int}$, the quantum energy, $E_{q}$, and the trap energy, $E_{tr}$, are given respectively by

\begin{equation}  
    E_{kin}(t) = \int \frac{\hbar^{2}}{2m} \left( \sqrt{\rho({\bf x}, t)}{\bf v}({\bf x}, t)\right)^{2} d^{2}\mathbf{r},
\label{eqn:KE}
\end{equation}

\begin{equation}  
      E_{int}(t) = \int g \left(\rho({\bf x}, t)\right)^{2} d^{2}\mathbf{r},
\label{eqn:IE}
\end{equation} 

\begin{equation}  
      E_{q}(t) = \int  \frac{\hbar^{2}}{2m}\left(\nabla \sqrt{\rho ({\bf x}, t)}\right)^{2} d^{2}\mathbf{r},
\label{eqn:QE}
\end{equation}

\begin{equation}  
      E_{tr}(t) = \int \rho({\bf x}, t)V_{tr}(\mathbf{x})  d^{2}\mathbf{r}.
\label{eqn:TrE}
\end{equation} 
Furthermore, we decompose the kinetic energy, $E_{kin}$, into a part due to the sound field, $E_{sound}$, and a part due to vortices, $E_{vortex}$ \citep{Kobayashi}.

\begin{equation}  
E_{kin} = E_{sound}+E_{vortex} 
\label{eqn:KSVE}
\end{equation}
At a given time t, the vortex energy, $E_{vortex}$, is obtained by relaxing the Gross-Pitaevskii equation in imaginary time, which yields the lowest energy state for a given vortex configuration
 \citep{Parker}. The sound energy is then recovered from $E_{sound} = E_{kin}-E_{vortex}$.

\bigskip
It is convenient to rewrite the Gross-Pitaevskii equation in dimensionless form in terms of the harmonic oscillator energy $\hbar\omega_{\perp}$, the harmonic oscillator length $\sqrt{\hbar/(2m\omega_{\perp})}$ and the harmonic oscillator time $\omega^{-1}_{\perp}$. We obtain:

\begin{equation}
 i\frac{\partial\psi}{\partial t}=\left [-\frac{1}{2}\nabla^{2}+V_{tr}+C|\psi|^{2}-\mu-\Omega L_{z}\right ]\psi,
\label{eqn:2D_GPE}
\end{equation}
where

\begin{equation}
V_{tr}(x,y)=\frac{1}{2}\left( x^{2}+y^{2}\right),
\label{eqn:2D_TP}
\end{equation}

\begin{equation}
C=\frac{4\pi Na}{L},
\label{eqn:2D_TP}
\end{equation}
N is the number of atoms and L the extension of the condensate in the z - direction. Unless stated otherwise, we set $C$ = 1400 in our calculations. In dimensionless units the initial estimate for $\mu$ is $\mu= 1/2 \sqrt{4C/\pi}$. The dimensionless form of the energy contributors are:

\begin{equation}  
    E_{kin}(t) = \int \frac{1}{2} \left( \sqrt{\rho({\bf x}, t)}{\bf v}({\bf x}, t)\right)^{2} d^{2}\mathbf{r},
\label{eqn:KE1}
\end{equation}

\begin{equation}  
      E_{int}(t) = \int C \left(\rho({\bf x}, t)\right)^{2} d^{2}\mathbf{r},
\label{eqn:IE1}
\end{equation} 

\begin{equation}  
      E_{q}(t) = \int  \frac{1}{2}\left(\nabla \sqrt{\rho ({\bf x}, t)}\right)^{2} d^{2}\mathbf{r},
\label{eqn:QE1}
\end{equation}
and

\begin{equation}  
      E_{tr}(t) = \int \rho({\bf x}, t)V_{tr}(\mathbf{x})  d^{2}\mathbf{r}.
\label{eqn:TrE}
\end{equation}
The calculation is performed in a square box of size D. We choose D so that it is larger than the trapped condensate, and impose boundary conditions $\psi$ = 0 at $x$ = $\pm D/2$, $y$ = $\pm D/2$. Typically, D = 10. The time stepping is a semi-implicit the Crank-Nicholson algorithm. 

\section{Results}
\begin{figure}[p]
\centering \epsfig{figure=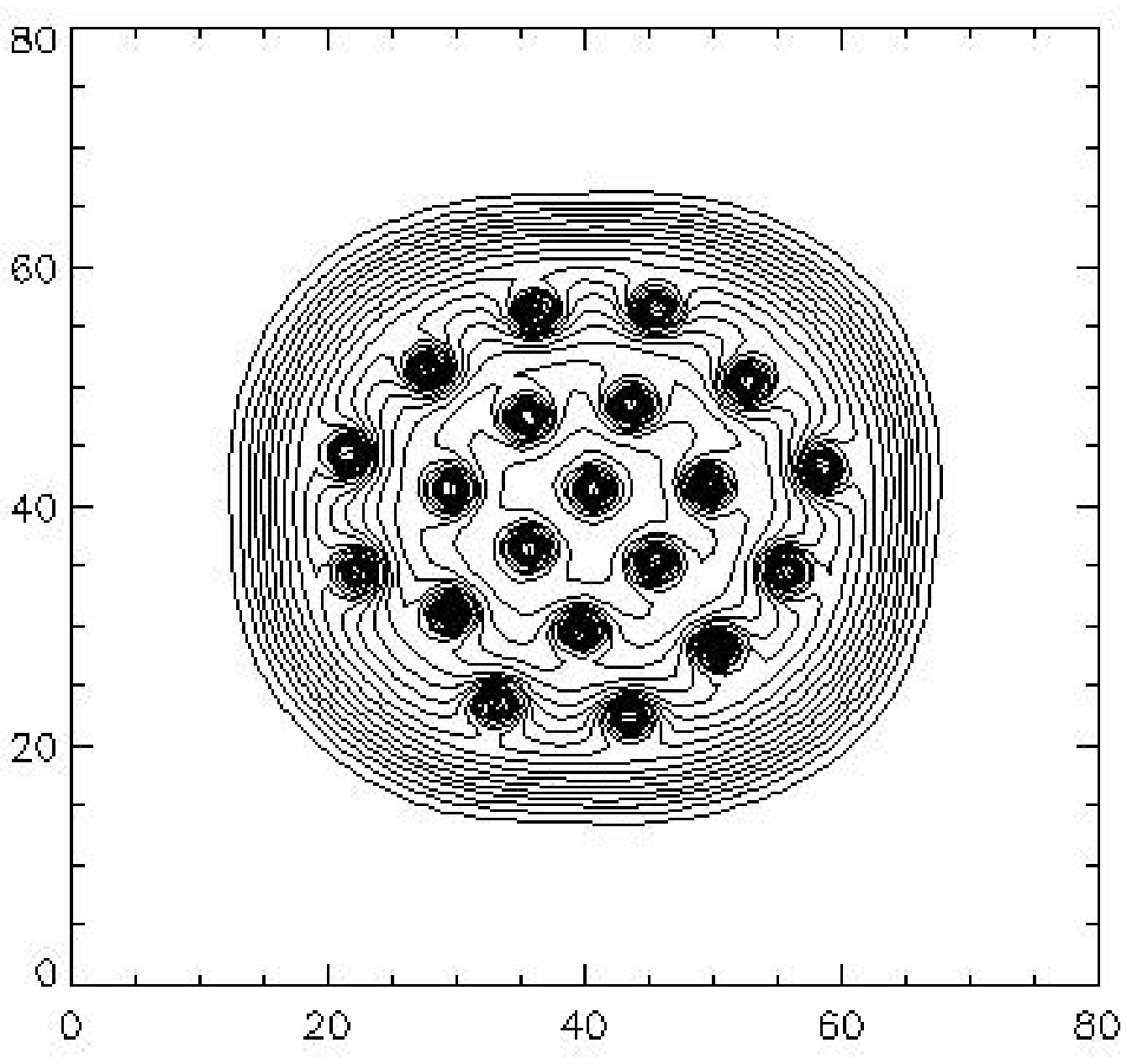,height=4in,angle=0,scale=0.6} 
\centering \epsfig{figure=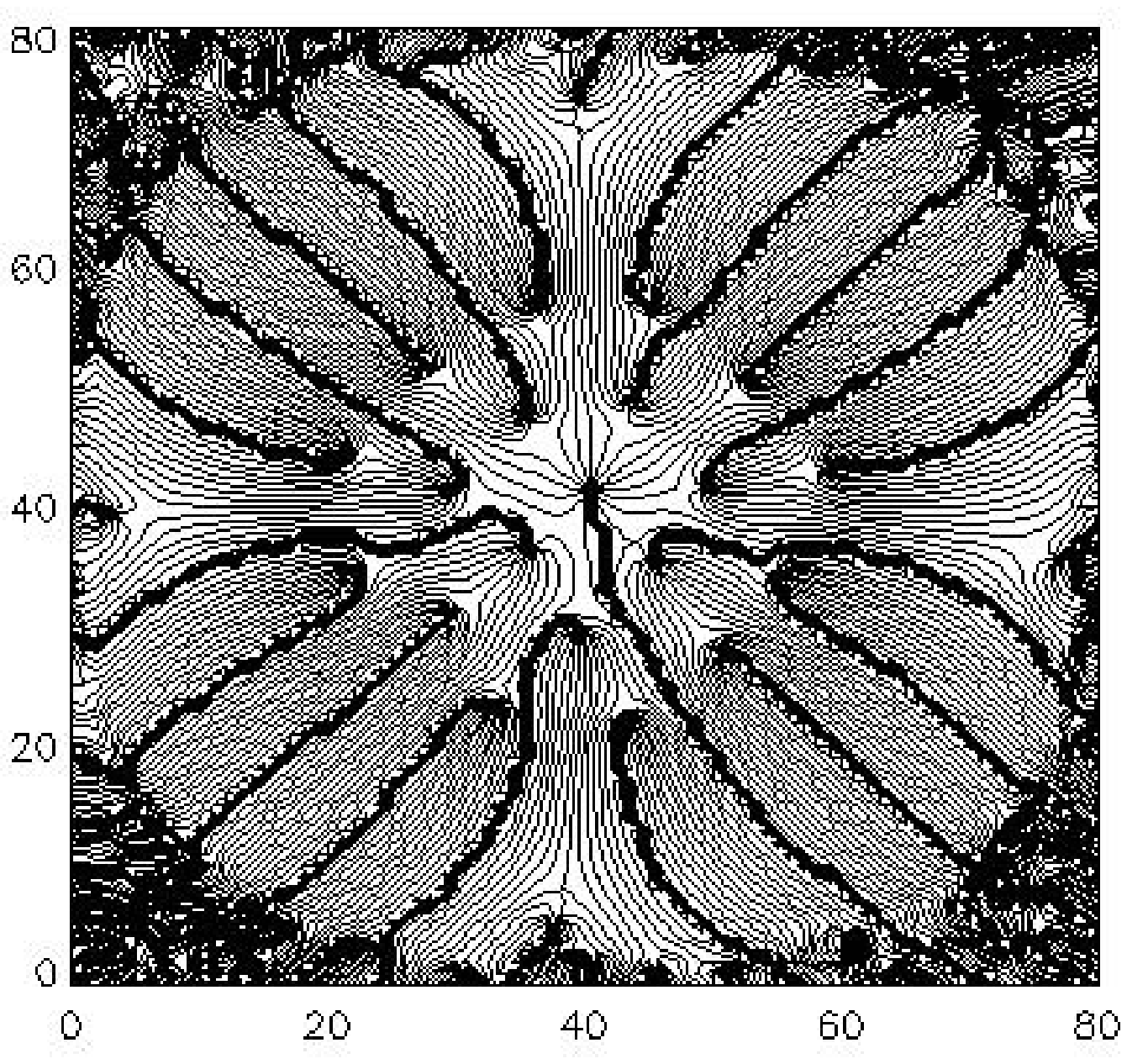,height=4in,angle=0,scale=0.6}
      \caption{Stable condensate with N = 20 vortices for $\Omega$ = 0.85 at t = 199.8 before the phase imprinting. The vortices are visible as holes in the density profile (left) or as discontinuities from $0$ to $2\pi$ of the phase profile (right).}
\label{fig:f1}
 \end{figure}

\begin{figure}[p]
\centering \epsfig{figure=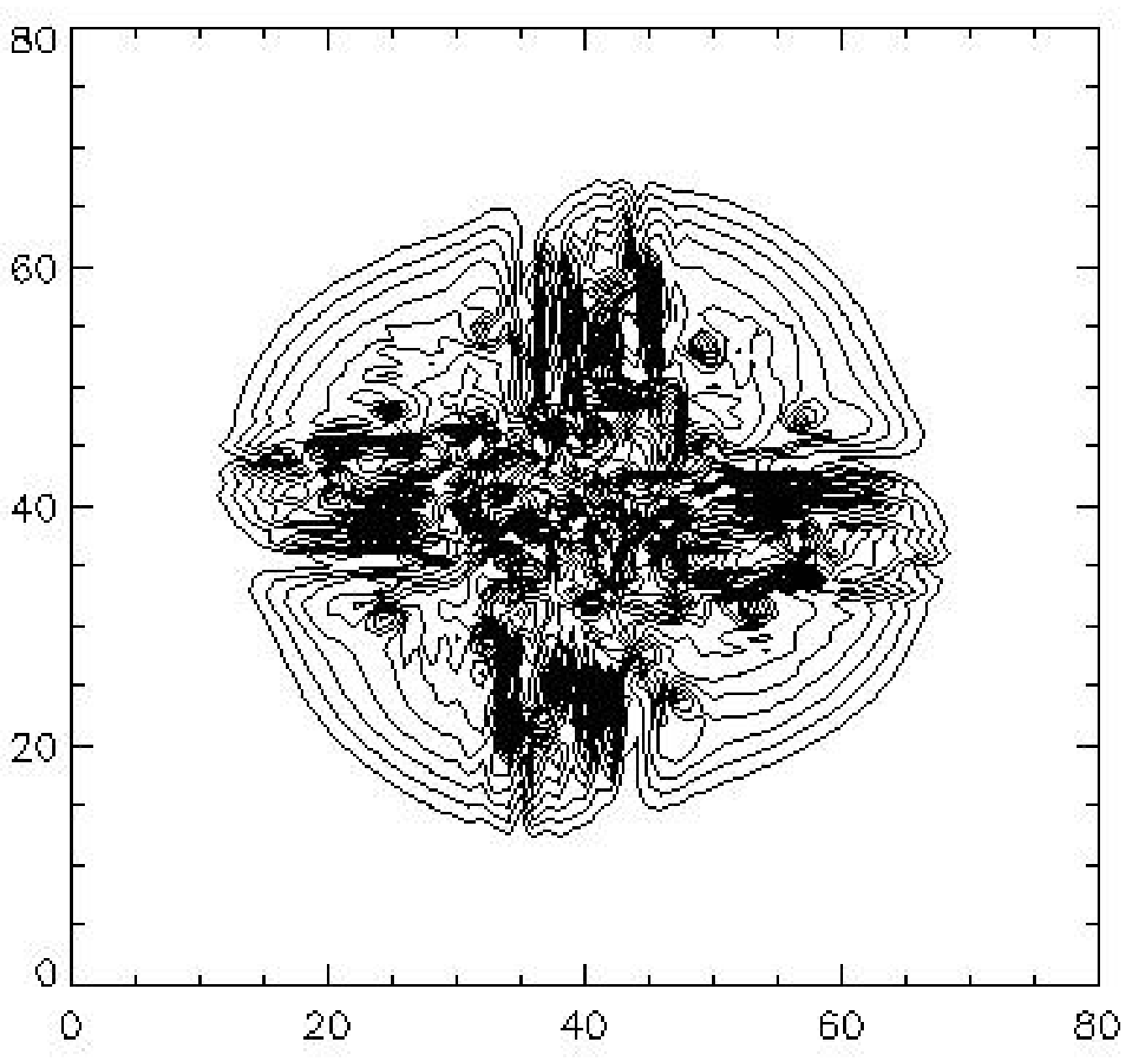,height=4in,angle=0,scale=0.6} 
\centering \epsfig{figure=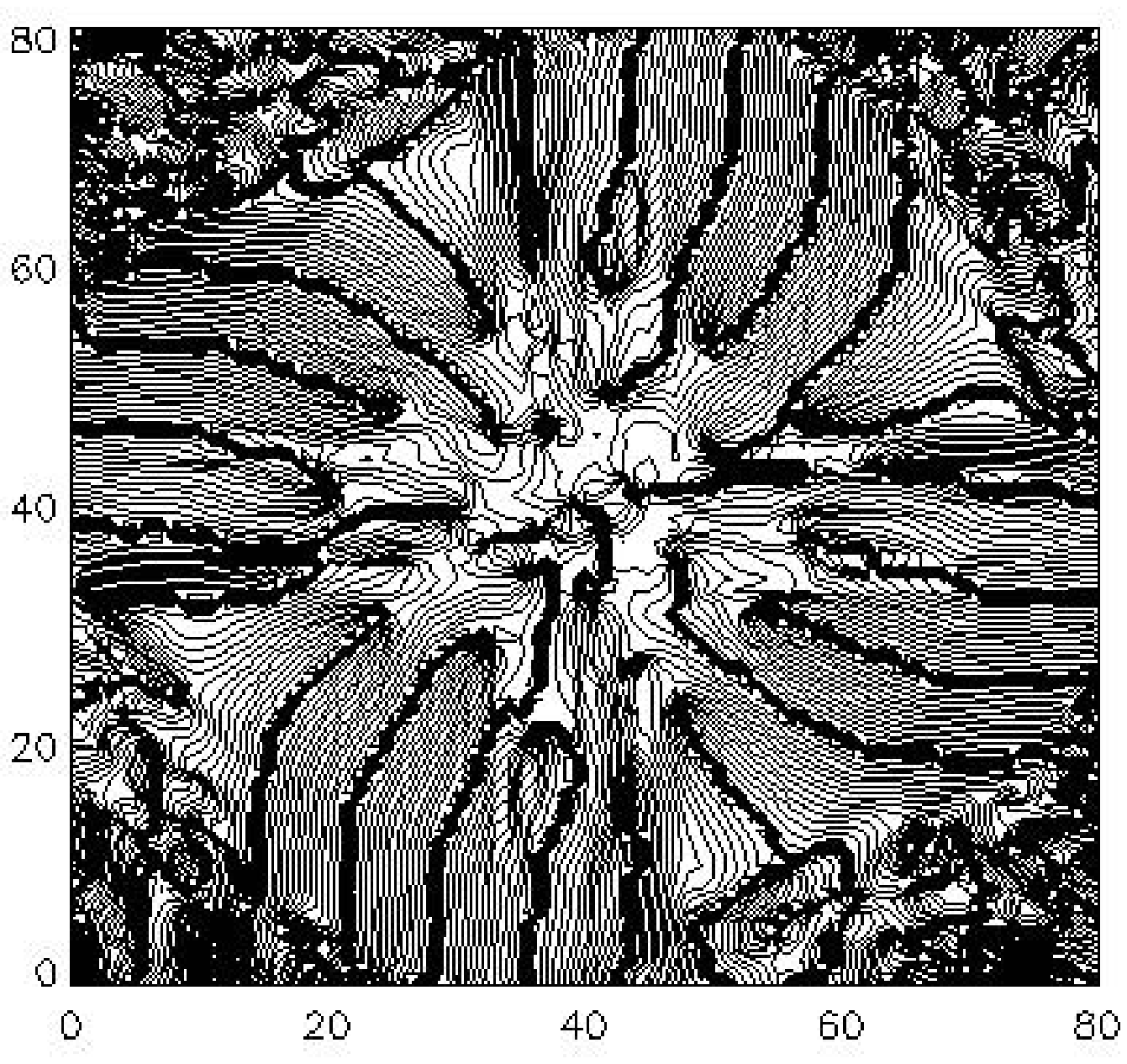,height=4in,angle=0,scale=0.6}
      \caption{Density (left) and phase (right) profiles at t = 200.4.}
\label{fig:f2}
 \end{figure}

Following  \citet{Leadbeater}, starting from a non-rotating condensate at time t = 0, we set $\Omega$ = 0.85 and create a stable lattice of 20 vortices in a rotating frame. This state mimics the solid body rotation of a rotating classical Euler fluid. The relaxation of the vortex system into a stable lattice is helped by the addition of a tiny dissipative term, $-\gamma \hbar \frac{\partial \Psi}{\partial t}$, with $\gamma$ $\ll$ 1 at the left hand side of Eq.~\ref{eqn:GP} and Eq.~\ref{eqn:2D_GPE}, as described by \citet{Tsubota}; this term models the interaction of the condensate with the surrounding thermal cloud at finite temperatures as described by \citet{Madarassy}. Density and phase profiles of the condensate are shown in Fig.~\ref{fig:f1}. At t = 200, we suddenly perform the following phase imprinting:

\begin{equation}  
\psi \to \psi \hspace{5mm} for \hspace{5mm} x<0,y<0 \hspace{5mm} and \hspace{5mm} x>0,y>0,
\label{eqn:PIM_IIa}
\end{equation}

\begin{equation}  
\psi \to e^{i\pi}\psi \hspace{5mm} for \hspace{5mm} x<0,y>0 \hspace{5mm} and \hspace{5mm}x>0,y<0. 
\label{eqn:PIM_IIb}
\end{equation}
For a large system with many vortices (as our system), the phase imprinting denotes that about half of quantized vortices change their sense of rotation. Fig.~\ref{fig:f2} shows the condensate just after the phase imprinting. Note that, the original condensate is divided into four equal parts, two parts rotating clockwise and two anticlockwise. This implies that the kinetic energy of the system is suddenly lowered. Hereafter, we shall consider the contributions to the energy for $t > 200$, after the phase imprinting.

\begin{figure}[p]
\centering \epsfig{figure=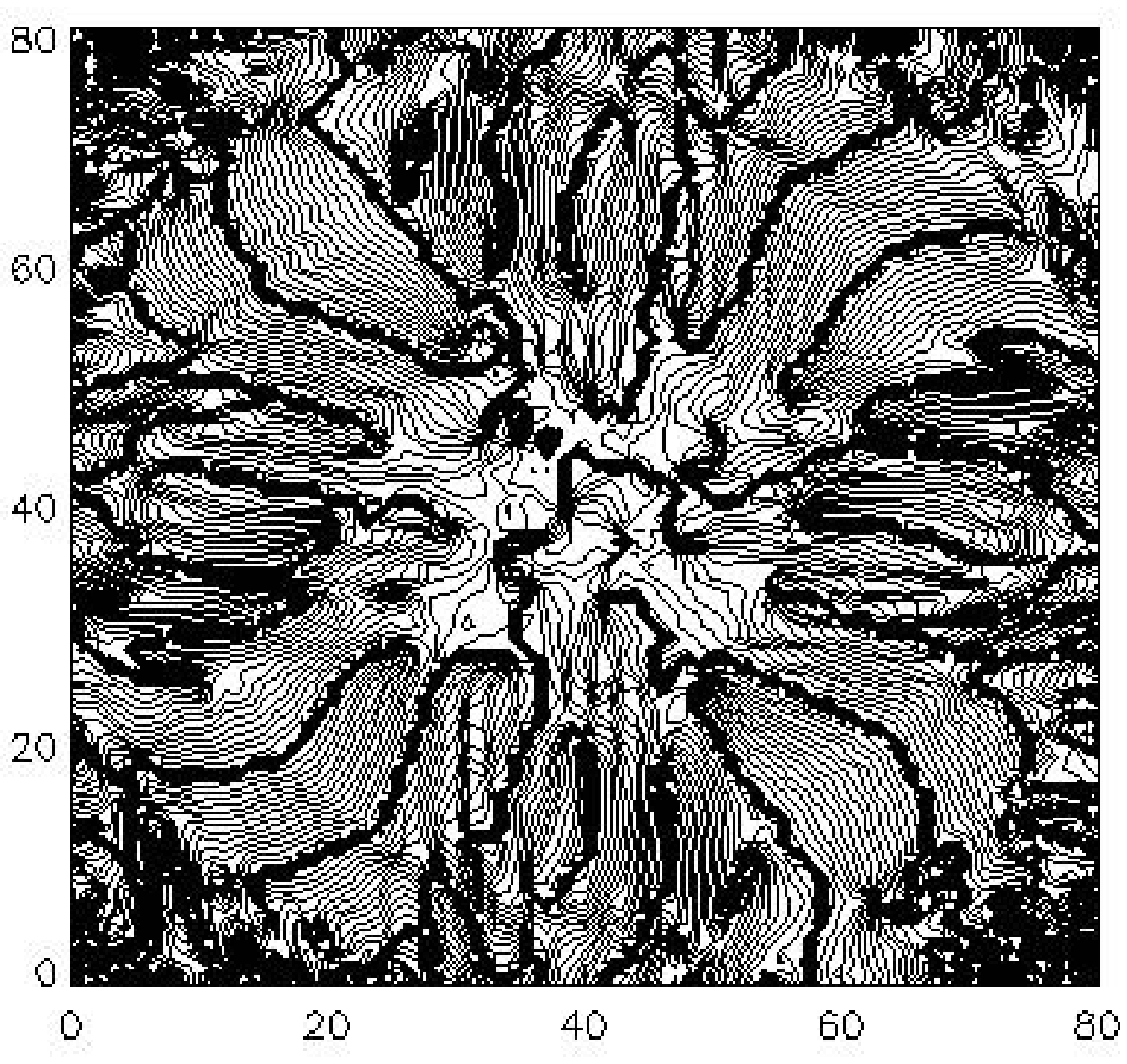,height=4in,angle=0,scale=0.6} 
\centering \epsfig{figure=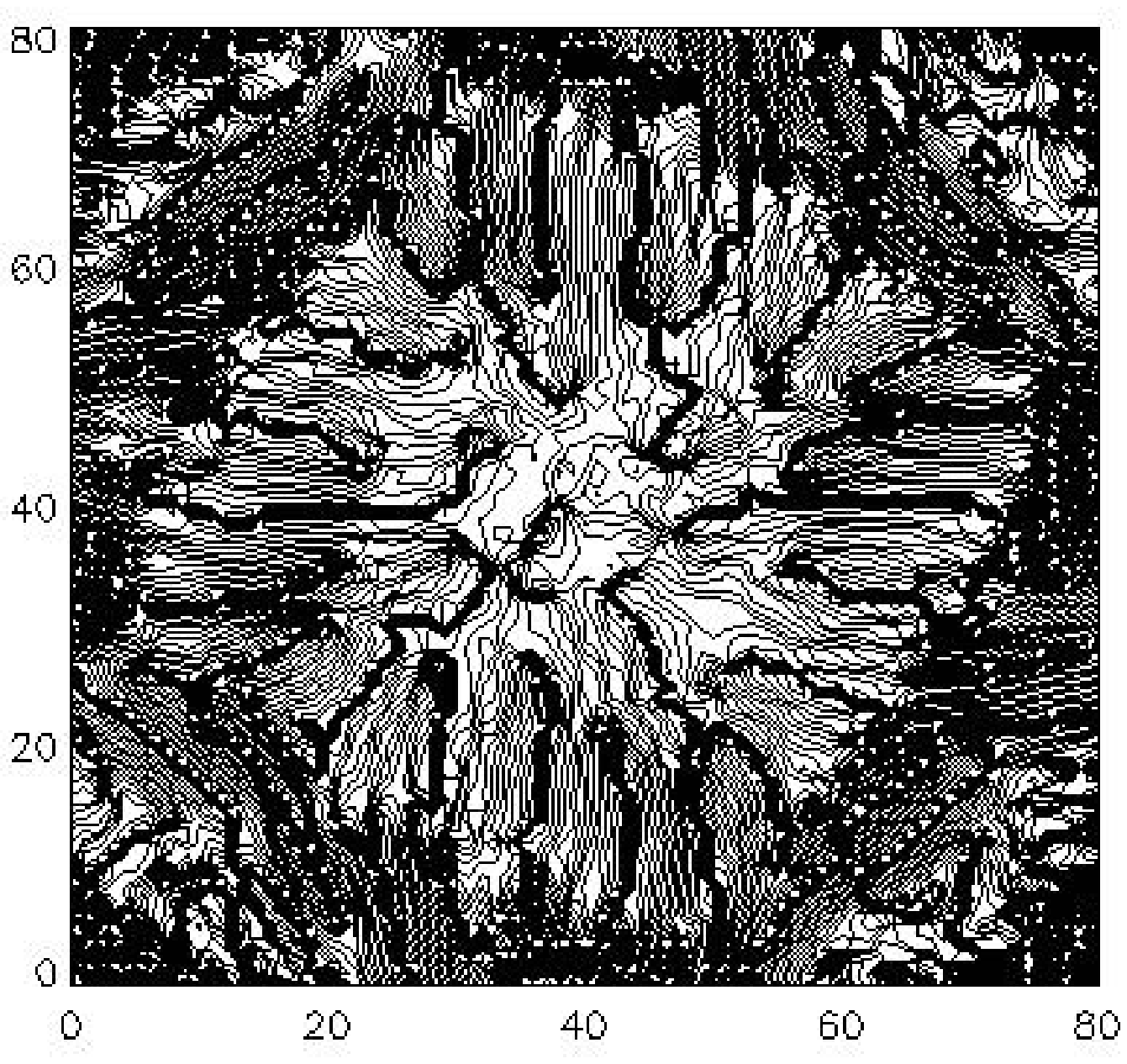,height=4in,angle=0,scale=0.6}
      \caption{Phase profile at t = 200.65 (left) and 200.9 (right).}
\label{fig:f3}
\end{figure}

\begin{figure}[p]
\centering \epsfig{figure=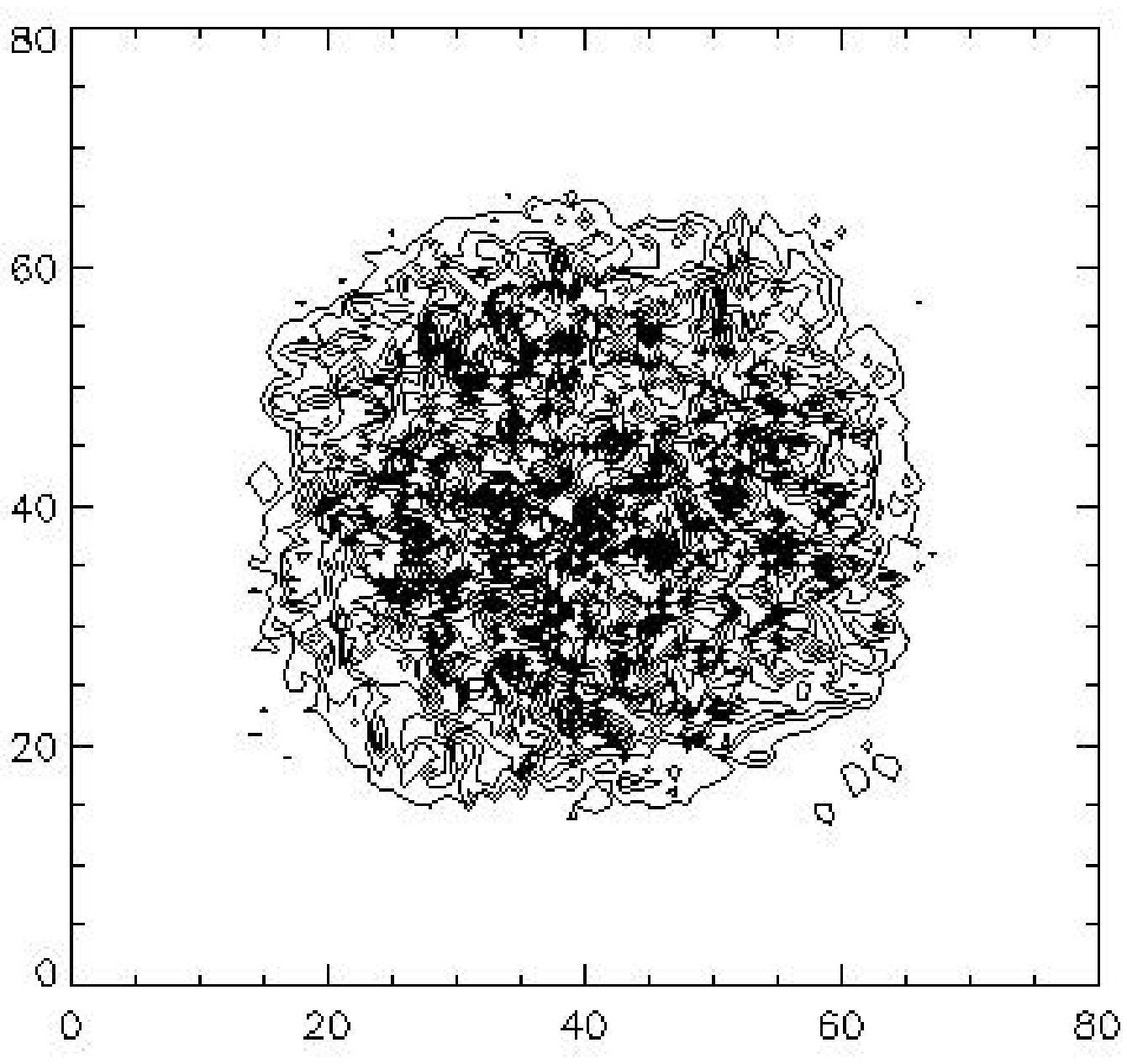,height=4in,angle=0,scale=0.6} 
\centering \epsfig{figure=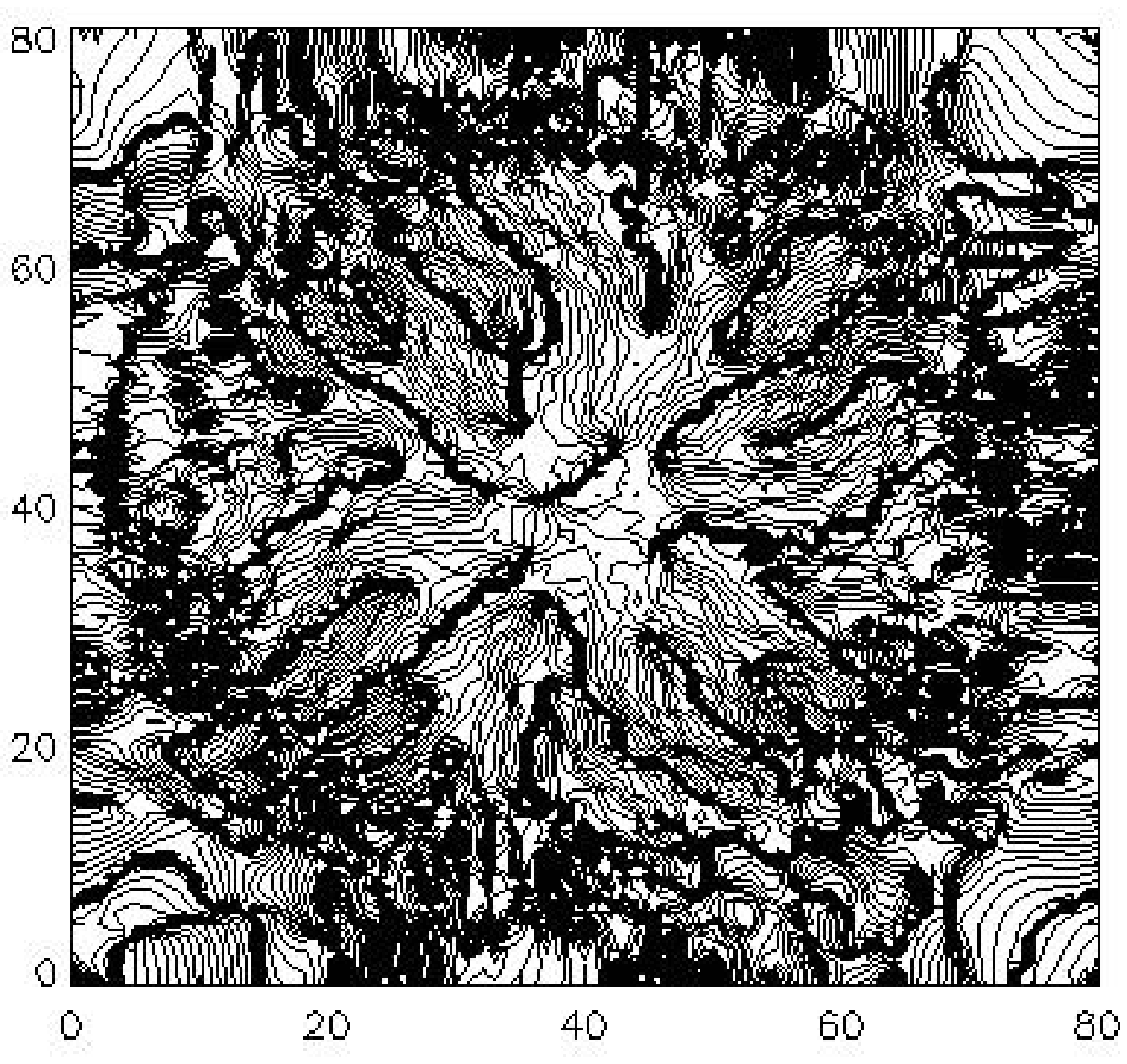,height=4in,angle=0,scale=0.6}
      \caption{Density profile (left) and phase profile (right) at t = 201.95.}
\label{fig:f4}
\end{figure}

\bigskip
After the phase imprinting, positive and negative vortices interact and move due the velocity field which they induce on each other. Fig.~\ref{fig:f3} shows phase profiles at t = 200.65 and t = 200.9. Fig.~\ref{fig:f4} shows density and phase profiles at t = 201.95. Fig.~\ref{fig:f4}a reveals that  large density oscillations (sound waves) are present in the condensate.

\begin{figure}[p]
\centering \epsfig{figure=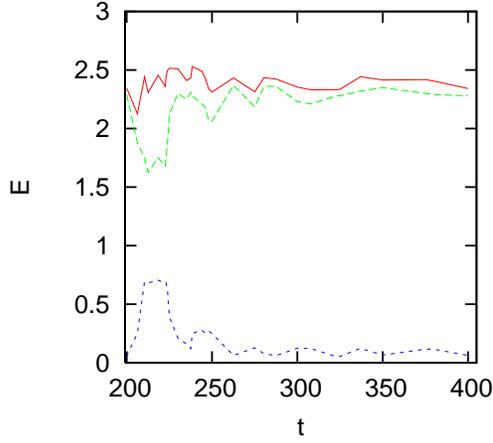,height=4in,angle=0,scale=0.6} 
\caption{Kinetic energy (solid curve, top), vortex energy (dashed curve, middle) and sound energy (dotted curve, bottom).}
\label{fig:f5}
\end{figure}

\begin{figure}[p]
\centering \epsfig{figure=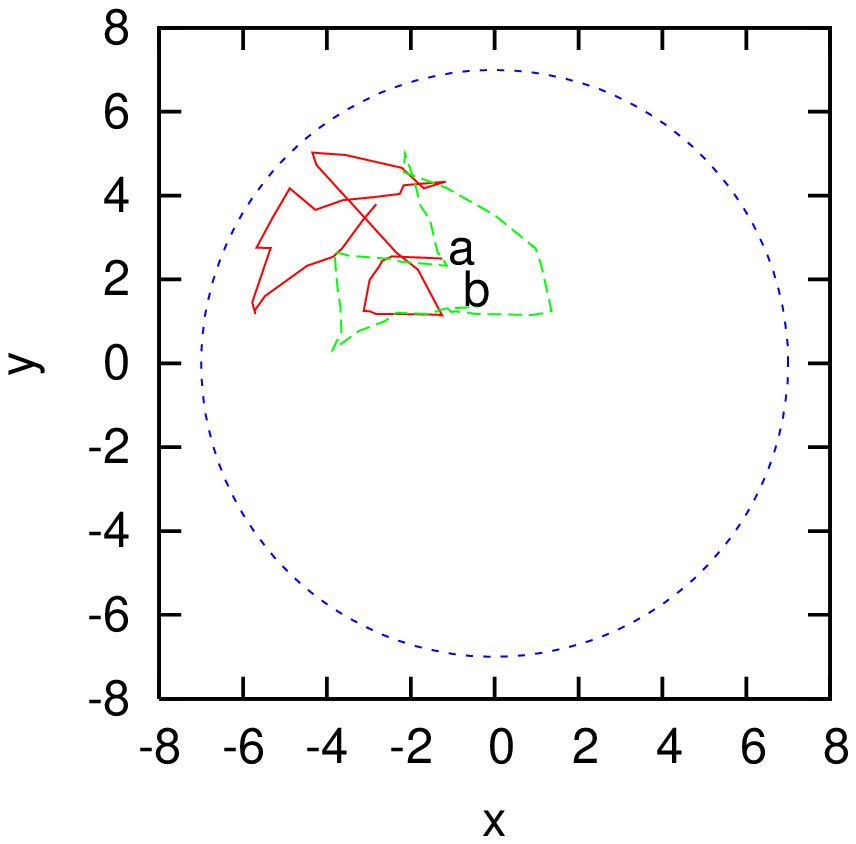,height=5in,angle=-90,scale=0.6}
\centering \epsfig{figure=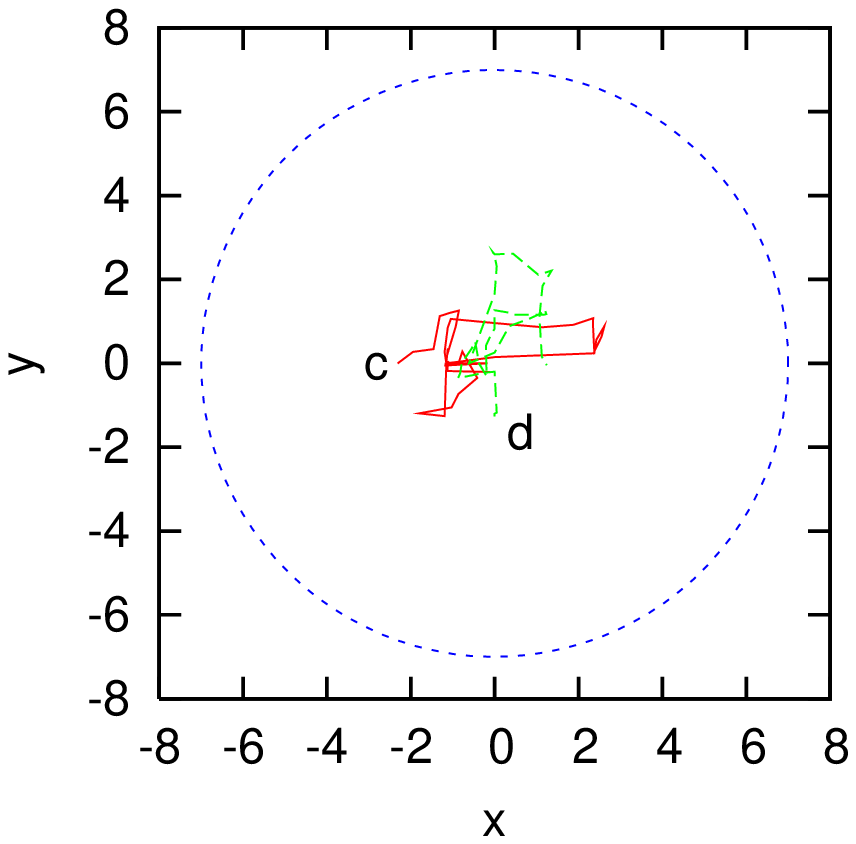,height=5in,angle=-90,scale=0.6} 
\caption{Left: Trajectories of positive vortex $a$ (solid curve) initially located at x = -1.25, y = 2.5 and of negative vortex $b$ (dashed curve) initially located at x = -1.12, y = 1.31. Right: Trajectories of positive vortex $c$ (solid curve) initially located at x = -2.31, y =  0 and of positive vortex $d$ (dashed curve) initially located at x = 0, y = -1.27, from t = 200.5 to t = 204.7.}
\label{fig:f6}
\end{figure}

\begin{figure}[p]
\centering \epsfig{figure=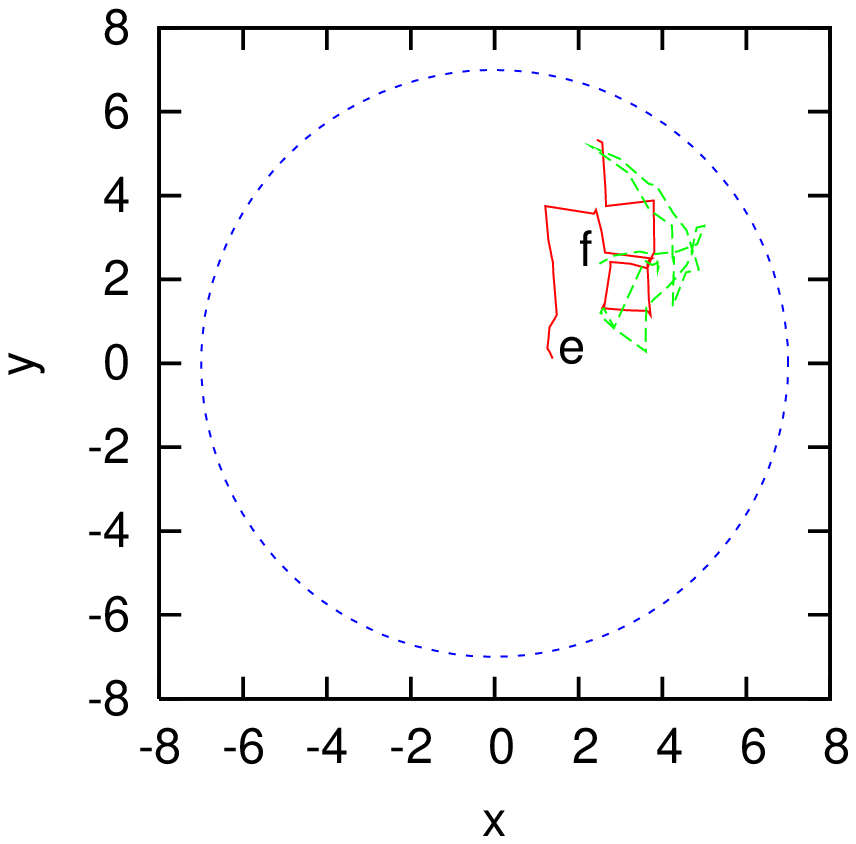,height=5in,angle=-90,scale=0.6}
\centering \epsfig{figure=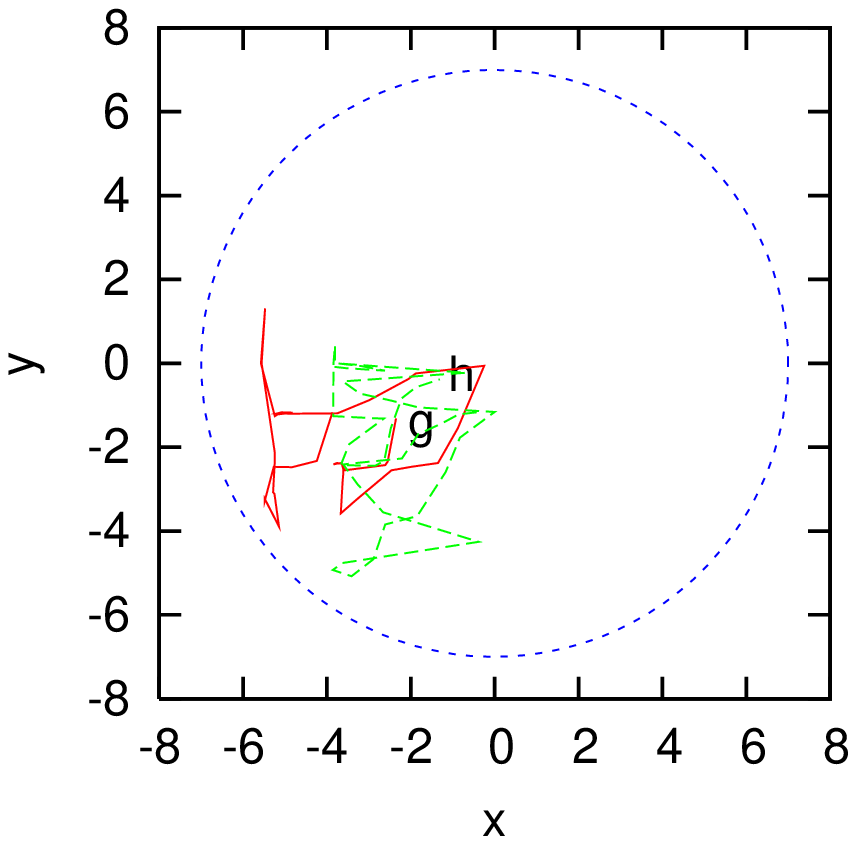,height=5in,angle=-90,scale=0.6} 
\caption{Left: Trajectories of positive vortex $e$ (solid curve) initially located at x = 1.38 , y = 0.12 and of positive vortex $f$ (dashed curve) initially located at x = 3.88, y = 2.45. Right: Trajectories of positive vortex $g$ (solid curve) initially located at x = -2.35, y =  -1.32 and of negative vortex $h$ (dashed curve) initially located at x = -0.43, y = -1.17, from t = 200.5 to t = 204.7.}
\label{fig:f7}
\end{figure}

\begin{figure}[p]
\centering \epsfig{figure=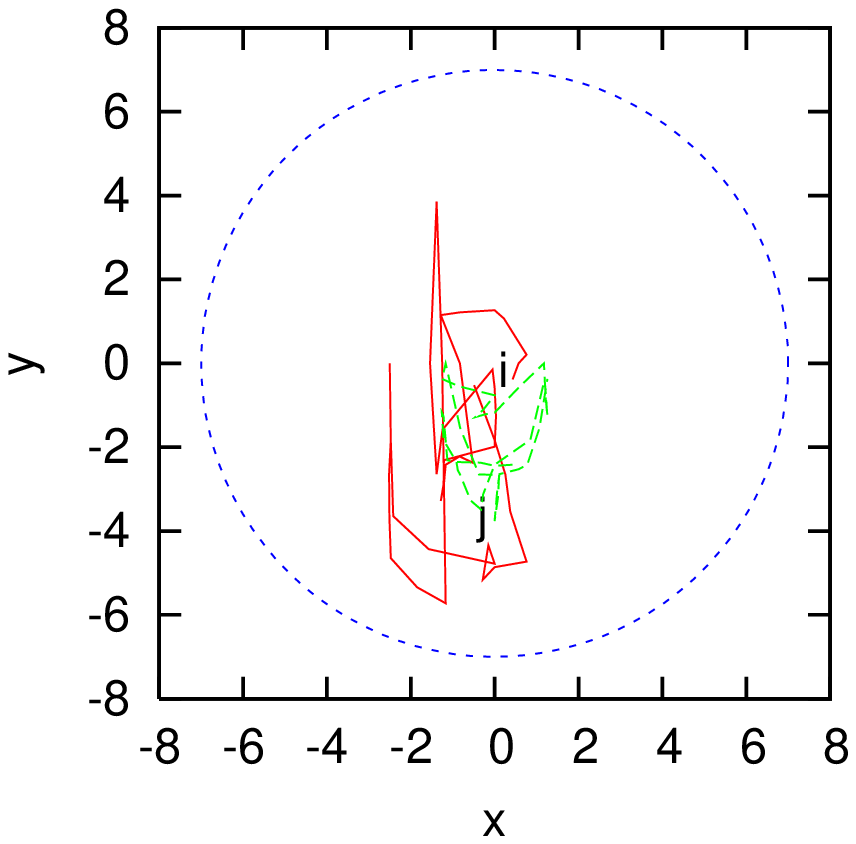,height=5in,angle=-90,scale=0.6}
\centering \epsfig{figure=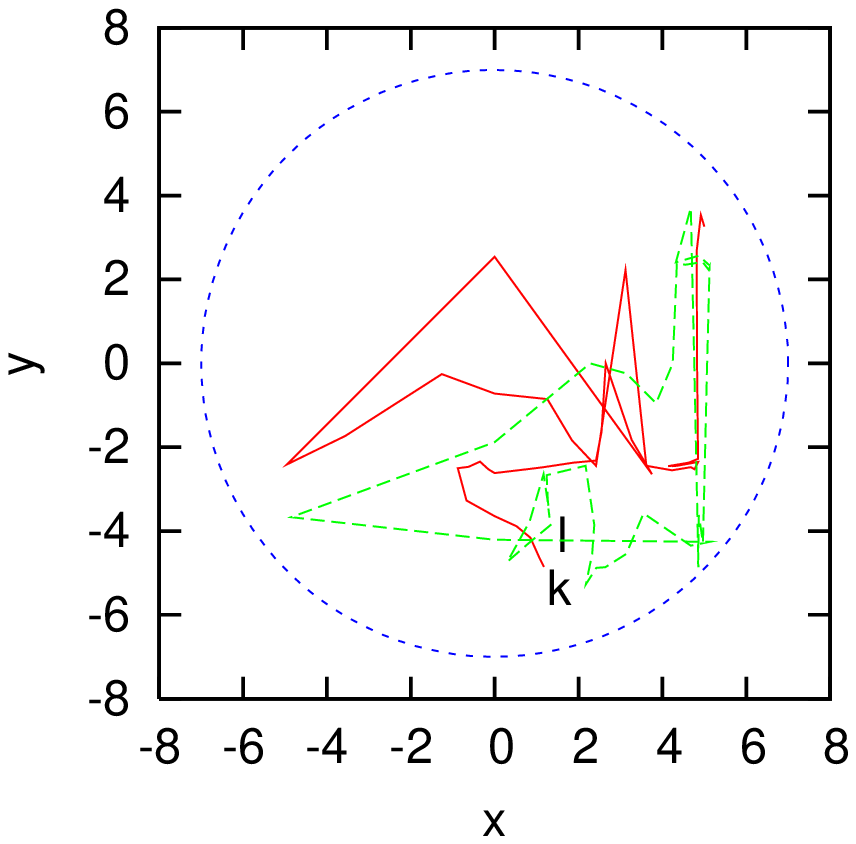,height=5in,angle=-90,scale=0.6} 
\caption{Left: Trajectories of positive vortex $i$ (solid curve) initially located at x = 0.43 , y = -0.38 and of negative vortex $j$ (dashed curve) initially located at x = 0.42, y = -2.42. Right: Trajectories of positive vortex $k$ (solid curve) initially located at x = 1.18, y =  -4.85 and of negative vortex $l$ (dashed curve) initially located at x = 1.35, y = -3.84, from t = 200.5 to t = 204.7.}
\label{fig:f8}
\end{figure}

\begin{figure}[p]
\centering \epsfig{figure=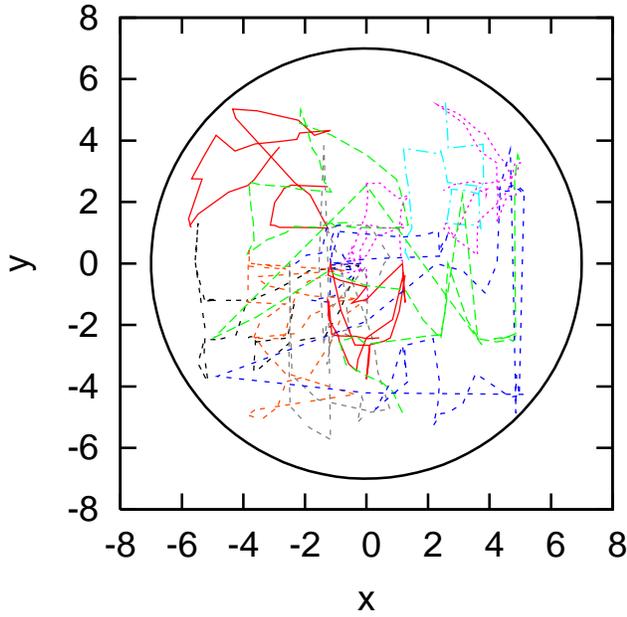,height=8in,angle=0,scale=0.6,angle=-90} 
\caption{Vortex trajectories of vortices $a,$ $b,$ $c,$ $d,$ $e,$ $f,$ $g,$ $h,$ $i,$ $j,$ $k,$  and $l,$ for times: 200.5 $\leq$ t $\leq$ 204.7.}
\label{fig:f9}
\end{figure}

\bigskip
Due to the confined and finite size of the condensate, the sound waves are reflected from the edge of the condensate and reinteract with the vortex pairs. We can see that when $E_{sound}$ is approaching its maximum, $E_{vortex}$ is approaching its minimum, see Fig.~\ref{fig:f5}. The interaction of positive and negative vortices leads to a disordered motion of the vortices. The trajectories of the vortices can be obtained by tracking the phase profiles in time. For clarity, we show vortex trajectories of two vortices at the time in Fig.~\ref{fig:f6}, Fig.~\ref{fig:f7} and Fig.~\ref{fig:f8}. Fig.~\ref{fig:f9} shows all these trajectories together.

\section{Discussion}
Our calculation shows that the method which we suggest can be used to create a system of positive and negative vortices in a disk-shaped condensate and that the vortices will move in irregular way under each other's influence. The method can thus be used to create a two-dimensional turbulent system, similar to the classical Onsager vortex gas \citep{Wang}. Clearly, a disk-shaped condensate created in the laboratory is a three-dimensional system, but, provided that, the thickness of the disk is only few times the coherence  length, the resulting turbulence will be sufficiently two-dimensional, because Kelvin waves will not be able to propagate along vortices in the z-direction.

\begin{figure}[p]
\centering \epsfig{figure=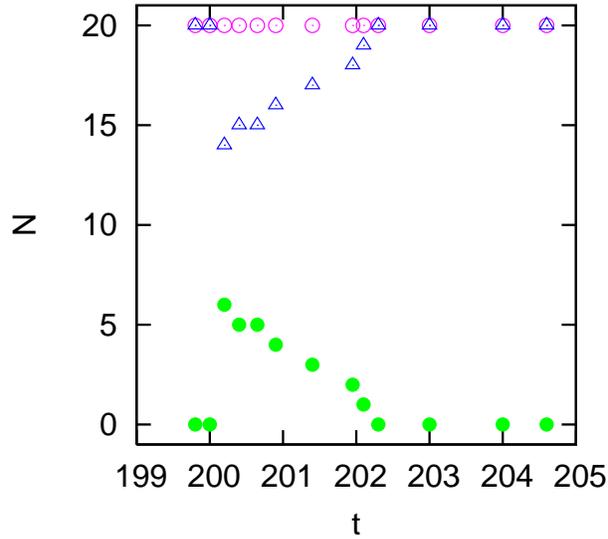,height=5in,angle=0,scale=0.6,angle=0} 
\caption{Number of total vortices, $N$ (circles), positive vortices, $N^{+}$ (triangles) and negative vortices, $N^{-}$ (filled circles) as a function of time if, after the phase imprinting at t = 200, $\Omega$ is not changed and remains $\Omega$ = 0.85.}
\label{fig:f10}
\end{figure}

\bigskip
It is useful to distinguish two possible evolutions after the phase imprinting:

\bigskip
(i): $\Omega$ and $\gamma$ are kept the same. After a transient turbulent state, the system will relax to the original vortex lattice as in Fig.~\ref{fig:f1} (essentialy repeating the original calculation but starting from a different initial condition). Negative vortices will move to the edge of the condensate and vanish, as shown in Fig.~\ref{fig:f10}.
 
\begin{figure}[p]
\centering \epsfig{figure=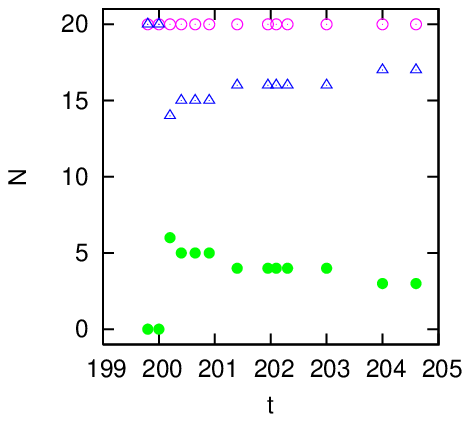,height=5in,angle=0,scale=0.6,angle=0} 
\caption{The number of total vortices, $N$ (circles), positive vortices, $N^{+}$ (triangles) and negative vortices, $N^{-}$ (filled circles) as a function of time if, after the phase imprinting at t = 200, $\Omega$ is set to $\Omega$ = 0.} 
\label{fig:f11}
\end{figure}

\bigskip
(ii): $\Omega$ is set to zero together with the dissipation, $\gamma$. If $\gamma \ne 0$, the system ought to eventually relax to the ground state with no vortices. If $\gamma=0$, the negative vortices remain in the condensate, as in Fig.~\ref{fig:f11}.

\bigskip
It is instructive to compare the number of total, negative and positive vortices in these two cases ($\Omega$ and $\gamma$ kept the same and $\Omega$ and $\gamma$ set to zero) for the same interval of time, between $t=199$ and $t=205$. We conclude, that in the case of (i), the negative vortices disappear quickly, but in the case of (ii), for that short time, the negative vortices stay in the condensate and the transient turbulent state lasts much longer. Over a much longer time scale, we expect that the condensate will become vortex-free, due to collisions and mutual annihilation of vortices with anti vortices.

\bigskip
There are many possible variations of the method. For example, we can obtain positive and negative vortices by setting: $\psi \to \psi$ for $y>0$ and $\psi \to \psi^{*}$ for $y<0,$ (where the star denotes the complex conjugate). If we set: $\psi \to \psi$ for $y>0$ and $\psi \to e^{i\pi}\psi$ for $y<0$ a dark soliton-like disturbance \citep{Jackson} is created along the x-axis. In a two-dimensional system this structure is unstable to the snake instability \citep{Feder} and vortex - anti vortex pairs (or vortex rings in three dimensions \citep{Mamaev} are created.
 
\section{Acknowledgements}
The authors are grateful to B. Jackson (deceased) and A. J. Youd for useful discussions.



\begin{thebibliography}{99}
\bibitem[\protect\citeauthoryear{Barenghi {\itshape{et al.,}}}{2001}]{Barenghi}
Barenghi, C. F., Donnelly, R. J. and Vinen, W. F., Quantized Vortex Dynamics and Superfluid Turbulence. {\itshape
(Springer-Verlag, Berlin),} 2001. 
\bibitem[\protect\citeauthoryear{Barenghi,}{2008}]{Barenghi1}
Barenghi, C. F., to be published in Physica D (2008).
\bibitem[\protect\citeauthoryear{Maurer and Tabeling,}{1998}]{Maurer}
Maurer, J., and Tabeling, P., Local investigation of superfluid turbulence. {\itshape
Europhys. Lett.,} 1998, {\bfseries 43}, 29.
\bibitem[\protect\citeauthoryear{Nore,}{1997}]{Nore}
Nore, C., Abid, M. and Brachet, M. E., Kolmogorov Turbulence in Low-Temperature Superflows. {\itshape
Phys. Rev. Lett.,} 1997, {\bfseries 78}, 3896.
\bibitem[\protect\citeauthoryear{Araki {\itshape{et al.,}}}{2002}]{Araki}
Araki, T., Tsubota, M. and Nemirovskii,S. K., Energy Spectrum of Superfluid Turbulence with No Normal-Fluid Component. {\itshape
Phys. Rev. Lett.,} 2002, {\bfseries 89}, 145301.
\bibitem[\protect\citeauthoryear{Kobayashi and Tsubota,}{2005}]{Kobayashi}
Kobayashi, M. and Tsubota, M., Kolmogorov Spectrum of Superfluid Turbulence: Numerical Analysis of the Gross-Pitaevskii Equation with a Small-Scale Dissipation. {\itshape
Phys. Rev. Lett.,} 2005, {\bfseries 94}, 065302.
\bibitem[\protect\citeauthoryear{Smith {\itshape{et al.,}}}{1993}]{Smith}
Smith, M. R., Donnelly, R. J., Goldenfeld, N. and Vinen, W. F., Decay of vorticity in homogeneous turbulence. {\itshape
Phys. Rev. Lett.,} 1993, {\bfseries 71}, 2583.
\bibitem[\protect\citeauthoryear{Kobayashi and Tsubota,}{2007}]{Kobayashi1}
Kobayashi, M. and Tsubota, M.,  Quantum turbulence in a trapped Bose-Einstein condensate. {\itshape
Phys. Rev. A,} 2007, {\bfseries 76}, 045603.
\bibitem[\protect\citeauthoryear{Berloff}{2004}]{Berloff}
Berloff, N. G., Interactions of vortices with rarefaction solitary waves in a Bose-Einstein condensate and their role in the decay of superfluid turbulence. {\itshape
Phys. Rev. A,} 2004, {\bfseries 69}, 053601.
\bibitem[\protect\citeauthoryear{Laval {\itshape{et al.,}}}{2004}]{Laval}
Laval, J.-P., Dubrulle, B. and Nazarenko, S. V., Fast numerical simulations of 2D turbulence using a dynamic model for subgrid motions. {\itshape
J.of Comp. Phys.,} 2004, {\bfseries 196}, 184-207.
\bibitem[\protect\citeauthoryear{Leadbeater {\itshape{et al.,}}}{2001}]{Leadbeater}
Leadbeater, M., Winiecki, T., Samuels, D. S., Barenghi, C. F. and Adams, C. S.,Sound Emission due to Superfluid Vortex Reconnections. {\itshape
Phys. Rev. A,} 2001, {\bfseries 86}, 1410.
\bibitem[\protect\citeauthoryear{Parker {\itshape{et al.,}}}{2004}]{Parker1}
Parker, N. G., Proukakis, N. P., Barenghi C. F. and Adams, C. S., Controlled Vortex-Sound Interactions in Atomic Bose-Einstein Condensates . {\itshape
Phys. Rev. Lett.,} 2004, {\bfseries 92}, 160403-1.
\bibitem[\protect\citeauthoryear{Burger {\itshape{et al.,}}}{1999}]{Burger}
Burger, S., Bongs,  K., Dettmer, S., Ertmer, W. and Sengstock, K., Dark Solitons in Bose-Einstein Condensates. {\itshape
Phys. Rev. Lett.,} 1999, {\bfseries 83}, 5198.
\bibitem[\protect\citeauthoryear{Dobrek {\itshape{et al.,}}}{1999}]{Dobrek}
Dobrek, Ł., Gajda, M., Lewenstein, M., Sengstock, K., Birkl, G. and Ertmer, W., Optical generation of vortices in trapped Bose-Einstein condensates. {\itshape
Phys. Rev. A,} 1999, {\bfseries 60}, R3381.
\bibitem[\protect\citeauthoryear{Tsubota {\itshape{et al.,}}}{2002}]{Tsubota}
Tsubota, M., Kasamatsu, K., and Ueda, M., Vortex lattice formation in a rotating Bose-Einstein condensate. {\itshape
Phys. Rev. A,} 2002, {\bfseries 65}, 023603.
\bibitem[\protect\citeauthoryear{Madarassy and Barenghi,}{2008}]{Madarassy}
Madarassy E. J. M., and Barenghi C. F., Vortex dynamics in trapped Bose-Einstein condensate. {\itshape
published in JLTP,} (2008).
\bibitem[\protect\citeauthoryear{Wang {\itshape{et al.,}}}{1999}]{Wang}
Wang, S., Sergeev, Y. A., Barenghi, C. F., and Harrison, M. A., Two-particle separation in the point vortex gas model of superfluid turbulence. {\itshape
J. Low Temp. Physics,} 2007, {\bfseries 149}, 65.
\bibitem[\protect\citeauthoryear{Pitaevskii and Stringari,}{2002}]{Stringari}
Pitaevskii, L. P. and Stringari, S.,Bose-Einstein Condensation. {\itshape
Clarendon Press, Oxford,} 2003, 
\bibitem[\protect\citeauthoryear{Parker and Adams,}{2005}]{Parker}
Parker, N. G. and Adams, C. S.,  Emergence and Decay of Turbulence in Stirred Atomic Bose-Einstein Condensate. {\itshape
Phys. Rev. Lett.,} 2005, {\bfseries 95}, 145301.
\bibitem[\protect\citeauthoryear{Jackson {\itshape{et al.,}}}{2007}]{Jackson}
Jackson, B., Barenghi, C. F. and Proukakis, N. P., Matter wave solitons at finite temperatures. {\itshape
J. Low. Temp. Phys.,} 2007, {\bfseries 148}, 387-391.
\bibitem[\protect\citeauthoryear{Feder {\itshape{et al.,}}}{2000}]{Feder}
Feder, D. L., Pindzola, M. S., Collins, L. A., Schneider, B. I. and Clark, C. W., Dark-soliton states of Bose-Einstein condensates in anisotropic traps. {\itshape
Phys. Rev. A,} 2000, {\bfseries 62 }, 053606.
\bibitem[\protect\citeauthoryear{Mamaev {\itshape{et al.,}}}{1996}]{Mamaev}
Mamaev, A. V., Saffman, M., and Zozulya, A. A., Propagation of Dark Stripe Beams in Nonlinear Media: Snake Instability and Creation of Optical Vortices. {\itshape
Phys. Rev. Lett,} 1996, {\bfseries 76}, 2262.

\end{thebibliography}
\end{document}